\documentclass[twocolumn,english]{revtex4}
\usepackage{babel,amsmath,amssymb,dcolumn}
\usepackage{hhline}
\usepackage{multirow}
\usepackage[dvips]{graphics}

\def\agt{\mathrel{\raise.3ex\hbox{$>$}\mkern-14mu\lower0.6ex\hbox{$\sim$}}}
\def\alt{\mathrel{\raise.3ex\hbox{$<$}\mkern-14mu\lower0.6ex\hbox{$\sim$}}}

\def\be{\begin{eqnarray}}
\def\ee{\end{eqnarray}}
\def\bsubeq{\begin{subequations}}
\def\esubeq{\end{subequations}}

\begin{document}
\title{Scalar field propagation in braneworld black hole scenario obtained from Nash theorem}
\author{R. D. B. Fontana}
\email{rodrigof@fma.if.usp.br}
\affiliation{*Universidade Federal da Fronteira Sul, Av. Fernando Machado, 108E, Centro, Chapec\'o, SC 89802-112, Brazil}
\author{C. Maia}
\email{casmaia@fis.unb.br}
\author{M.D.Maia}
\email{maia@unb.br}
\affiliation{$\dag$ Instituto de F\'isica, Universidade de Bras\'ilia, Bras\'ilia, DF 70910-900, Brazil}
\author{S. S. A. Silva}
\email{ssilva@ita.br}
\affiliation{Instituto Tecnol\'ogico de Aeron\'autica, Pra\c ca Marechal Eduardo Gomes, 50, Vila das Acacias, São Jos\'e dos Campos, SP 12228-900, Brazil}

\begin{abstract}
We determine the scalar field evolution of a braneworld localized black hole with both dark matter and dark energy components, obtained within a dynamical and continuous embedding formalism by use of Nash's theorem. We further extract the associated quasinormal modes for both de Sitter-Schwarzschild-Dark matter and anti-de Sitter-Schwarzschild-Dark matter solutions, for the different causal structures (one, two or three horizon spacetimes) given by the space of parameters. By comparison with standard General Relativity solutions, we infer possible observable astrophysical differences, and remark on modifications to previous AdS/CFT correspondence scenarios.

\end{abstract}

\maketitle
\section{Introduction}
The advent of gravitational wave astronomy has opened up a unique window to test various extensions of General Relativity, among them the different braneworld scenarios proposed in the last two decades. To date, the joint observation of gravitational and electromagnetic signals by GW170817 and GRB170817 \cite{ligo} already provides for a comparison between the propagation of low frequency ($\leq$ 1 kHz) gravitational waves and gamma ray photons, setting constraints on the possible leaking of gravitational waves to any extra dimensions \cite{pardo}. 
\par Braneworld models may also inprint observable signatures on gravitational waves through tail effects present on its ringdown phase, even in scenarios where the asymptotic propagation of the signal is identical to light waves \cite{andriot,brustein,yunes}. The resolution of the tail of a gravitational wave is the more challenging part of our present day observations, but it also comprises a wider sample covering all binary compact systems, i.e. even when no electromagnetic signal is present, as in the case of black hole binary progenitors. The most direct way to compute such tail effects is by analysis of quasinormal modes, which we present here for a particular braneworld scenario.
\par The idea that our present 4-dimensional spacetime is a hypersurface in a 5-dimensional (or greater) bulk has been under analysis since the work of Randal-Sundrum (RS) \cite{rs,reviewbrane}, who described a braneworld scenario where matter and gauge interactions are restricted to our 4D brane while gravitational degrees of freedom are not. Even though the physical motivation and results of RS scenarios are robust, they so far lack any natural geometrical implementation
within the framework of General Relativity. In order to mesh a 4-dimensional spacetime
into higher dimensional ones, these models feature junction conditions, boundary terms and mirror
symmetries due to mostly \textit{ad hoc} assumptions taken in order to render analytically tractable models. 
\par In fact, the general problem of the embedding of a Riemannian manifold to a higher dimensional one remounts to the birth of differential geometry and is non-trivial.              After decades of slow progress, J. Nash was the first to show in
1956 \cite{nash} a universal recipe for the embedding of differentiable
manifolds. By use of Nash's theorem, one of the authors has been able to build braneworld scenarios where our observable universe is embedded in a bulk spacetime without use of arbitrary junction conditions and artificial symmetries. As of now, that program has been successful to produce cosmological models where the present acceleration of the universe has a natural geometrical origin \cite{maia}.
                \par The same natural immersion techniques of our formalism have been used to
obtain spherically symmetric compact solutions in braneworlds \cite{ira}, as part of a program to establish the range of parameters allowed by observations, i.e. solar system, galaxy and local clusters. In particular, assuming only a 5D constant curvature bulk spacetime, our immersion formalism leads to a generalization of the standard Schwarzschild-de Sitter solution (SdS). Our particular interest in this solution is doubly motivated by (i) the surprising fitting it provides for galaxy rotation curves, connecting dark matter and dark energy by means of a unified geometrical origin \cite{ira}; and (ii) its connection to AdS/CFT conjecture and its holographic analogues \cite{horohub}. So in order to better understand and classify our solution in comparison to SdS, we here obtain the characteristic quasinormal modes of this braneworld scenario, which is also a necessary step to obtain the ringdown signature of their gravitational emission, to be treated in a separate work. 
\par Our paper is divided as follows. In Section \ref{secbbh} we review our embedding formalism and its results, in Section \ref{sscalar} and \ref{snummet} we review the scalar field and numerical evolution methods, which are applied to the de-Sitter-Schwarzchild case in Section \ref{ssdsdm} and the anti-de Sitter-Schwarzschild one in Section \ref{ssdsddm}. We conclude with final comments on Section \ref{conc}.

%

\section{Braneworld black hole \label{secbbh}}

We briefly review our braneworld formalism \cite{maia}. Given the $n$-dimensional background manifold $V_{n}$ to be isometrically embedded in $V_{D}$, $D > n$, by a map $M: V_{4} \to V_{n}$ with the $D$-dimensional bulk basis $X^{A}$ taking values at the $n$-dimensional coordinates $x^{\nu}$, i.e. $X^{A} = X^{A}(x^{1}, x^{2}, ... x^{n})$, we define

\be
\label{form1}
G_{AB}X_{,\mu}^{A}X^{B}_{,\nu} = g_{\mu \nu}, \\
G_{AB}X_{,\mu}^{A}N^{B}_{,a} = 0, \\
G_{AB}N^{A}_{,a}N^{B}_{,b} = g_{ab} = \pm 1 ,
\ee
where $G_{AB}$ is the bulk metric, $g_{\mu\nu}$ the brane metric and $N^{A}$ the $D-n$ vectors orthonormal to the brane. We set the bulk as a dynamical spacetime in functionally dependent on the braneworld geometry, foregoing any artificially static or rigid embedding. By use of Nash's theorem  \cite{nash}, it is possible to show the perturbation and evolution of the brane remains isometrically embedded \cite{maia} by means of the generalized York relation:
\be
\label{york}
K_{\mu\nu a} = -\frac{1}{2}\frac{\partial g_{\mu\nu}}{\partial \xi^{a}},
\ee
where $K_{\mu\nu a}$ is the extrinsic curvature and $\xi^{a}$ the small parameters variation along the extra dimensions. Applying this formalism for the embedding of a spherically symmetric brane to a 5D bulk of constant curvature $k$,

\be
\label{form2}
R_{ABCD} = k(G_{AC}G_{BD} - G_{AD}G_{BC}),
\ee
and by use of the Gauss-Codazzi equations
\be
\label{form3}
R_{\alpha\beta\gamma\delta} = \pm (K_{\alpha\gamma}K_{\beta\delta} - K_{\alpha\delta}K_{\beta\gamma}) \\+ \nonumber R_{ABCD}X^{A}_{,\alpha}X^{B}_{,\beta}X^{C}_{,\gamma}X^{D}_{,\delta};
\ee

\be
\label{form4}
K_{\alpha[\beta;\gamma]} = R_{ABCD}X^{A}_{,\alpha}N^{B}X^{C}_{,\beta}X^{D}_{,\gamma}; \\
\ee
with positive sign for de Sitter (dS) and negative sign for anti-de Sitter (AdS) bulk, we arrive at the braneworld metric \cite{ira}

\be
\label{ex4}
g_{\mu \nu} = (-a(r), b(r),r^2,r^2\sin^2 \theta), 
\ee
with
\be
\label{b1}
a(r) \equiv a_\pm = b_\pm^{-1}= 1 - \frac{S}{r}  \pm (cr + k)^2.
\ee
Depending on the positive (dS) or negative (AdS) signal we shall have 3, 2 or 1 horizon acording to the chosen parameters. Compared to  solutions of standard General Relativity, the extra terms in the function $a_\pm$ arise from the influence of the intrinsic curvature of the bulk projected onto the brane: they represent a black hole in an AdS/dS-universe parametrized by $c$ and a dark matter-type component of parameter $k$ \cite{ira}.

The causal structure of the solution is determined by the horizon spheres with the equation $a(r)=0$. This equation leads us to at least one real root, although it may be negative depending on the range of parameters.  Establishing the three roots of the equation as $r_1, r_2$ and $r_3$ and assuming positive mass $S>0$, we treat the different causal structures separately.

\section{Scalar field in a fixed geometry\label{sscalar}}

 The propagation of a scalar field in a fixed geometry follows the Klein-Gordon equation defined as
\be
\label{a1}
\Box \Phi = \frac{1}{\sqrt{-g}}\partial_\mu[\sqrt{-g} g^{\mu \nu} \partial_\nu \Phi] = 0.
\ee
In general we may not suppose the geometry remains the same for any field perturbation unless the field decays in time, its contribution being a second order effect. This will be the case for stable quasi-normal modes as defined  in the next sections. 

In a spherically symmetric solution for the Einstein  Equations, we can decompose the field in angular, radial and temporal parts. As a consequence, the angular part can be expressed as spherical harmonics with eigenvalues $l(l+1)$:
\be
\nonumber
\Phi (r,t, \theta, \phi) = R(r) T(t)Y_l^m(\theta, \phi) \rightarrow \Box^{(ang)} \Phi = -l(l+1) \Phi.
\\
\label{a2}
\ee
Assuming now a diagonal metric as in eq. (\ref{ex4}), we define the tortoise coordinate system given by
\be
\label{a3}
dr_* = \sqrt{\frac{b}{a}}dr,
\ee
in order to avoid the singularity throughout integration of the field equation: the event horizon is  then placed at $-\infty$, and radial infinity at $+\infty$. In the presence of a cosmological horizon, we shall assume this to be the spatial infinity. 

We can obtain a simpler wave equation, rescaling the field as $\Phi \rightarrow \frac{\Psi}{r}$ so the Klein-Gordon equation is written as
\be
\label{a5}
\left[ -\frac{\partial^2}{\partial t^2} + \frac{\partial^2}{\partial r_*^2}- V(r) \right] \Psi (r) = 0, 
\ee
and with a diagonal metric we have
\be
\label{a6}
V(r) = a\left [-\frac{a \partial_rb-b\partial_ra }{2rab^2} + \frac{l(l+1)}{r^2} \right].
\ee

For the Schwarzschild case, the first part of the potential reads $4M/r^3$, and  outside the event horizon it is always positive. This turns out to produce only stable quasi-normal modes for the solution, as the damping part remains negative \cite{horohub} (i. e., it yields a stable field that decays in time). 

\section{Numerical Methods\label{snummet}}

With the scalar field equation given by  eqs. (\ref{a5}) and (\ref{a6}), we have a number of different methods used to extract the field profile in time domain and the quasi-normal frequencies. 

In order to integrate the scalar field to obtain its evolution profile in time, we use two null coordinates $u$ and $v$, defined as functions of $r_*$ and $t$, $t=v+u$ and $r_*=v-u$. The scalar field equation reads
\be
\label{d1}
\left[ -4 \frac{\partial}{\partial u}\frac{\partial}{\partial v} - V(r) \right] \Psi (r) = 0.
\ee
Now with a discrete grid in these coordinates,
\be
\label{d2}
\frac{\partial}{\partial v} \Psi & = & \lim_{\Delta v \rightarrow \infty} \frac{\Delta \Psi}{\Delta v} \simeq  \frac{\Psi_E-\Psi_W}{\Delta v}  \\
\label{d3}
\frac{\partial}{\partial u} \Psi & = & \lim_{\Delta u \rightarrow \infty} \frac{\Delta \Psi}{\Delta u} \simeq  \frac{\Psi_N-\Psi_S}{\Delta u}
\ee
the equation for the scalar field reads
\be
\label{d4}
\Psi_N = \Psi_E + \Psi_W - \Psi_S + \Delta u \Delta v V(r) \frac{\Psi_W +\Psi_E}{8}. 
\ee
Given an initial condition for the field profile as a gaussian package,
\be
\label{d5}
\Psi (u_0 , v) = exp \left[ - \frac{(v-v_c)^2}{2\sigma^2}\right], \hspace{0.2cm} \Psi (u, v_0) = const.,
\ee 
we are in position to obtain the time evolution of the field and eventually analyze the stability of the spacetime to the field propagation. 

To obtain the quasi-resonances there are several different possibilities in the literature and we choose to work with mainly three: the Prony method \cite{rosa}, the JWKB method \cite{roiycli} and the Horowitz-Hubeny method \cite{horohub}. As those methods are abundantly described in several references elsewhere, we shall not describe them here, restricting ourselves to applying the first two methods for the dS black hole and the first and third methods for the AdS one.

\section{Schwarzschild-dS-dM Black Hole \label{ssdsdm}}

\begin{figure}[h]
\caption{Limit functions $U_\pm$ for  a two horizons causal structure dS black hole. The yellow region represents possible values for dark energy, $c$ for which $a_-=0$ produces three real roots.}
\resizebox{\linewidth}{!}{\includegraphics*{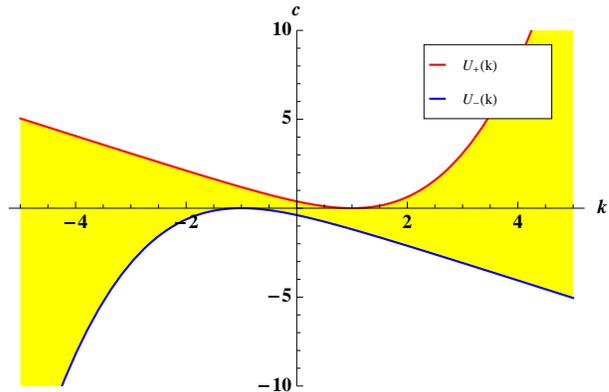}\label{f2}}
\end{figure}

Our de-Sitter-Schwarzschild-type black hole has a metric coefficient $a$ given by
\be
\nonumber
-g_{tt}=1 - \frac{S}{r} - (cr+k)^2=\frac{-c^2 \left( r^3 + \frac{2k}{c}r^2+ \frac{k^2-1}{c^2}r+\frac{S}{c^2} \right)}{r/c^2}
\\
\label{e22}
\ee
The causal structure is given by the possible horizons arising from $a=0$. Defining $r_1,r_2$ and $r_3$ as the three roots, then $r_1r_2r_3=-S<0$ and we must have at least one negative root. As a consequence, there are two different possibilities for the causal structure of this spacetime,
\vskip2ex
{\it (i) Two Horizons-spacetime}: shall be the case if the following condition holds,
\\
(D) $kc<0$ or $k^2<1$ when $kc>0$; 
\\
(E) $U_-<c<U_+$; 
\\ 
with $U_\pm= U_\pm (S,k)= \left( \frac{2k^3-18k}{27S} \pm \frac{2}{27S}\sqrt{( k^2+3 )^3}   \right)$. We must bring to attention the fact that condition (E) produces 3 real roots while condition (D) assures two of them to be positive. Figure \ref{f2} plots the functions $U_\pm$ with $S=1$ expressing the validity of three real roots solution in terms of cosmological constant (yellow region). Still if $k<-1$ ($k>1$), we must cut off the region $c<0$ ($c>0$) for in these regions all roots are negative.
\vskip2ex
{\it (ii) Naked Singularity}: shall be the case if the above conditions do not hold.
\\ 

The case without horizons will not be treated in the present work, as it lacks physical relevance given the presence of a naked singularity.

Once we fix the causal structure of the spacetime, we can choose a value for $S$ to integrate the scalar field. The qualitative behavior of its propagation depends on the geometry parameters and not on the chosen constants of the boundary conditions of the field. For the cases shown here, unless otherwise stated, we take $S=1$ without loss of generality.

\subsection{Effective Potential}

The potential for the scalar field propagation in the present case reads
\be
\nonumber
V(r ) = \left[ 1 - \frac{S}{r} - (cr+k)^2 \right] \left[ \frac{S}{r^3} - 2c^2 - \frac{2ck}{r}+\frac{l(l+1)}{r^2}  \right].
\\
\label{b2}
\ee  
\begin{figure}[h]
\caption{Potential for multiple $l$: $k=2S=-2c=2$. Similar to Schwarzschild-dS case, for $l=0$ we have $V_\mathbb{Y}<0$ between horizons, being still the case for $l=1$.}
\resizebox{\linewidth}{!}{\includegraphics*{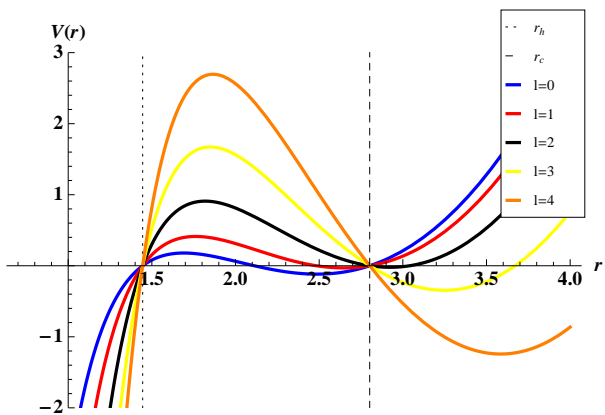}\label{f3}}
\end{figure}

We designate the region between horizons as $\mathbb{Y}$ and the potential for this range as $V_\mathbb{Y}$. A plot of this potential for the two horizons case, with given parameters, follows in figure \ref{f3}. 

As expected, the higher the multipole number, the higher the peak of the potential, happening $V_{\mathbb{Y}}>0$ from a critical value of $l$, depending on the black hole parameters. 

\begin{figure}[h]
\caption{Potential for high $k$: $k/100=S=-10c=1$, ($r_h \sim 990$). There is a negative region in the potential for different $l$, up to $l=13$.}
\resizebox{\linewidth}{!}{\includegraphics*{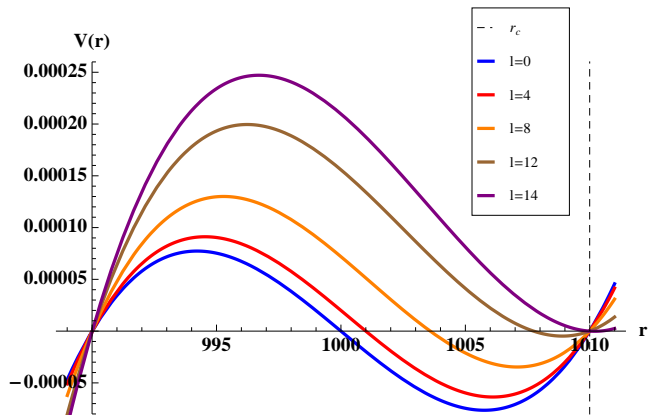}\label{f4}}
\end{figure}

For high values of $k$, being $|c|$ small, we have a greater number of multipoles for which $V_\mathbb{Y}<0$ at part of region $\mathbb{Y}$. In figure \ref{f4} we see a plot of the potential for different multipole numbers: we have $V_\mathbb{Y}>0$ throughout the entire $\mathbb{Y}$ only from $l=14$ on. If we take $|c|$ near it™s critical value (in this case, $-100<c<0$, $c_{crit}\sim -100$) we see a smaller number of multipoles for which $V_\mathbb{Y}<0$ in some $r$, but still greater than for the case of small $k$, e.g. for $c = -90$, and  $V_\mathbb{Y}<0$ occurs up to $l=7$.

Given the potential, we may analyze the field propagation over this geometry and extract the quasi-normal frequencies for œstable spacetimes. In the next subsection we treat the field propagation by integration of the Klein-Gordon equation in null coordinates.

\subsection{Scalar field time profile}

The acquisition of the field profiles follows the integration of a gaussian packet throughout Cauchy surfaces. The initial packet is of no importance in the field evolution in the quasinormal ringing phase and late time behavior: for every data of compact support \cite{koko} we will end up with the same field profile. In this sense, we can obtain some preliminary information about the stability of the field. If in time-domain the profile decays as a damped oscillator, or exponentially, or even goes to a constant value with time, we may consider it stable under perturbations. It is then possible the extraction via prony-method of the quasinormal frequencies \cite{rosa}. On the contrary, if by evaluation in time the field increases, we have unstable behavior suggesting the geometry must change. In this case the final stage of this geometry should be analyzed in a full non-linear formalism for gravitational perturbations.

\begin{figure}[h]
\caption{Time domain profiles with $l=0$. Here $S=-2k=1$ and the range for the cosmological term reads $-0.11 \simeq c \simeq 0.76$}
\resizebox{\linewidth}{!}{\includegraphics*{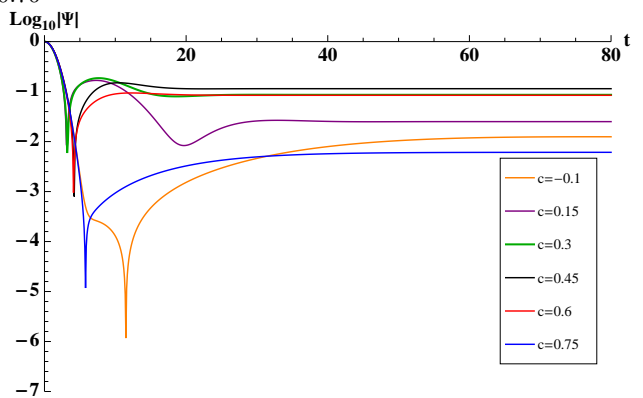} \label{f5}}
\end{figure}

In figure \ref{f5} we can see the time domain profile for $l=0$, the most negative of all potentials between horizons: no instabilities seems to occur for small values of $k$ and $c$; the fields evolve to a constant value after a small amount of time. The ringing phase is dumped before completing one oscillation for the displayed cases, which also occurs in all other $l=0$ cases here obtained. This is typical in dS-Schwarzschild black holes as well \cite{brachi, Molina}, for which the constant value of the field scales to the cosmological constant.

\begin{figure}[h]
\caption{Field profiles with $S=l=-10c/3=1$.}
\resizebox{\linewidth}{!}{\includegraphics*{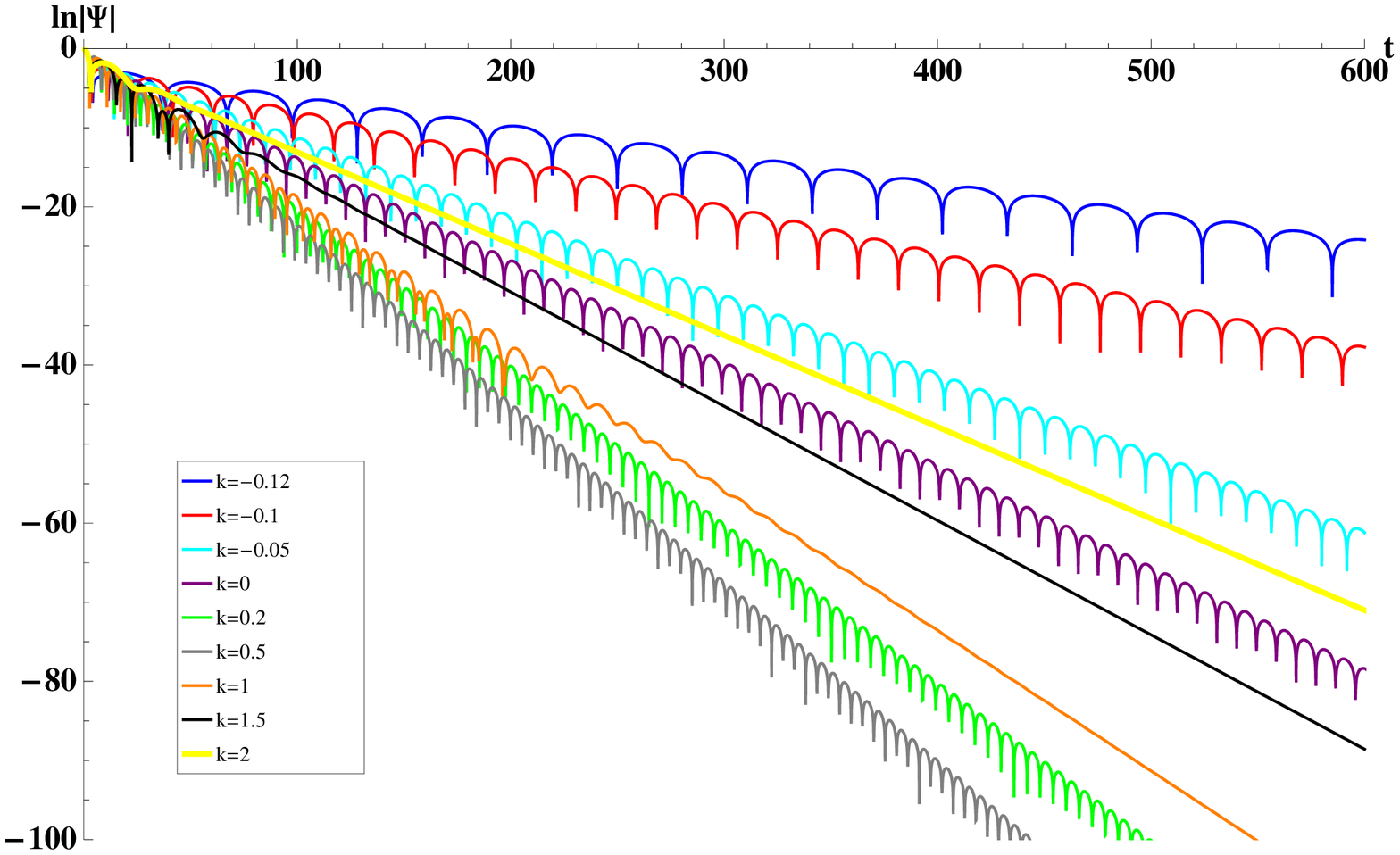}\label{f6}}
\end{figure}

For $l=1$, the evolution of a scalar field with different $k$ are shown in figure \ref{f6}.

\begin{figure}[h]
\caption{Field profiles with $S=l=-10c/3=1$.}
\resizebox{\linewidth}{!}{\includegraphics*{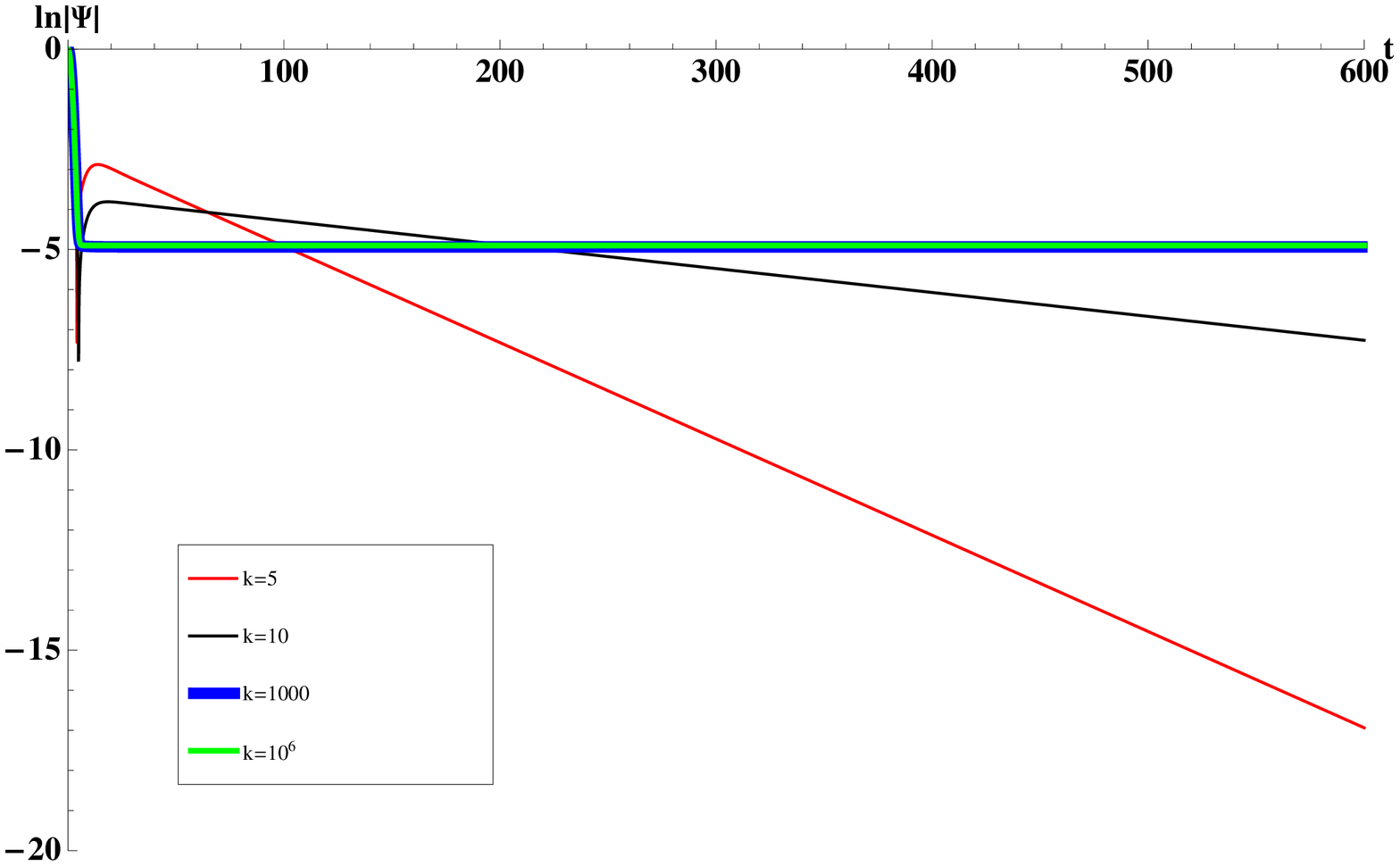}\label{f7}}
\end{figure}

Two different behaviours, related to the presence of a dark matter component in the field evolution, can be apprehended: (i) First, the ringing phase happens for a small range of $k$. With the used parameters, when $k>2$, the field rapidly turns to exponential decay. This is also present in other multipole numbers: the higher the $k$, in general, the smaller the ringing phase in time domain, and the smaller the coefficient of the exponential decay. In figure \ref{f7} we plot the field evolution, where we list different profiles for high $k$. 

(ii) Second, given $k$ for a horizon-encapsulated black hole $k_{crit}\sim -0.133$, the damping of the ringing phase increases up to the point where the phase fades away from the field evolution and exponential decay dominates. For $k>0$ we have field profiles with higher damping (in the quasinormal spectrum) and when $k<0$, on the contrary, the damping is smaller, when compared to the standard SdS black hole. 

\begin{figure}[h]
\caption{Field profiles with $S=-10c=10k=1$.}
\resizebox{\linewidth}{!}{\includegraphics*{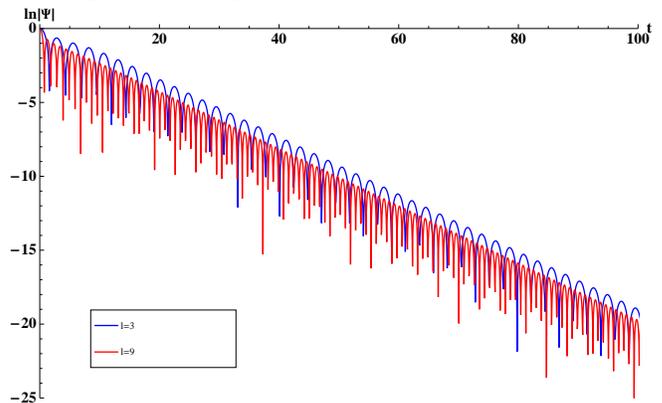}\label{f8}}
\end{figure}

\begin{figure}[h]
\caption{Field profiles with $S=-10c=k/80=1$.}
\resizebox{\linewidth}{!}{\includegraphics*{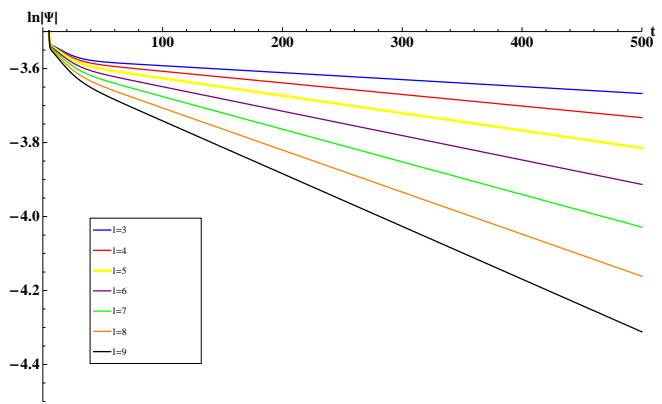}\label{f9}}
\end{figure}

Similar to the field behavior displayed above, the quasinormal spectrum for different $l$ appears only for small $k$. In figures \ref{f8} and \ref{f9} we see the field profiles for $k=0.1$ and $k=80$ respectively. In the second case, all field profiles rapidly evolve to an exponential decay, scaling up with the black hole parameters in cases of high $k$.

\subsection{Quasinormal Modes}

Given the field evolution in time domain, we are now in position to obtain the quasinormal modes with the pony method \cite{rosa} or its exponential decay with linear regression. Assuming a time evolution of type $\Psi (t) \rightarrow e^{-i \omega t}$, by use of the signals generated in the previous section we can extract the quasifrequencies $\omega = \omega_R - i \omega_I$ or the late time behavior, $\omega=-i\alpha$.  

We obtained a relatively small deviation between data collected via $6^{th}$-order JWKB and the one calculated via pony method, of the order of $10^{-4}$ (similar to \cite{sa1}). Comparing our results to \cite{sa1}, for example with $S=2$, $c=\sqrt{0.02/3}$ and $k=0$, the fundamental mode reads, $\omega_{l=1}^{JWKB} =0.2603-0.0911i$ and $\omega_{l=2}^{JWKB} =0.43461-0.08858i$, close to the resonance we obtained with the characteristic integration method, $\omega_{l=1}^{CI} =0.26076-0.09132i$ and $\omega_{l=2}^{CI} =0.43462-0.08857i$.

We are foremost interested in the effect introduced for the dark matter component in the SdS-like black hole. We summarize in table \ref{table:tab1} our results for the range of parameters in which a quasinormal spectrum is present, i.e. that of small $k$.
\begin{table}[ht]
\caption{Small $k$ cases.}
\label{table:tab1}
\centering
\begin{tabular}{|c|c|c|}
\hline
$k$ & Characteristic integration &  $6^{th}$-order JWKB  \\
\hline
$-0.12$ & $0.10316-0.036240i$ & $0.10317-0.036215i$ \\
\hline
$-0.1$ & $0.16626-0.059751i$ & $0.16621-0.059741i$ \\
\hline
$-0.05$ & $0.26660-0.099563i$ & $0.26654-0.099784i$ \\
\hline
$0$ & $0.33873-0.12918i$ & $0.33880-0.12941i$ \\
\hline
$0.1$ & $0.44368-0.17209i$ & $0.44379-0.17218$ \\
\hline
$0.2$ & $0.51441-0.20007i$ & $0.51444-0.20012i$ \\
\hline
$0.3$ & $0.55947-0.21690i$ & $0.55940-0.21693i$ \\
\hline
$0.4$ & $0.58361-0.22487i$ & $0.58347-0.22490i$ \\
\hline
$0.5$ & $0.59307-0.22681i$ & *** \\
\hline
$1$ & $0.46973-0.24777i$ & ***  \\
\hline
\end{tabular}
\end{table}
For values $k\gtrsim 0.5$ it is not possible to apply the JWKB method, given the presence of negative regions $V_\mathbb{Y}<0$ in between horizons.

For high values of $k$, we see in table \ref{table:tab2} the value of the field exponential decay coefficient.

\begin{table}[ht]
\caption{Large $k$ cases.}
\label{table:tab2}
\begin{tiny}
\begin{tabular}{|c|c|c|c|} 
\hline
Parameters & $\alpha$ & Parameters & $\alpha$ \\ 
$S=-10c=k/80=1$ & & $S=-10c/3=l=1$ & \\
\hline
$l=3$ & $0.0001877062$ & $k=1$ & $0.1774639$ \\ 
\hline
$l=4$ & $0.0003132374$ & $k=1.5$ & $0.1445301$ \\ 
\hline
$l=5$ & $0.0004705987$ & $k=2$ & $0.1158478$ \\ 
\hline
$l=6$ & $0.0006600949$ & $k=5$ & $0.02402919$ \\ 
\hline
$l=7$ & $0.0008820978$ & $k=10$ & $5.965834\cdot 10^{-3}$ \\ 
\hline
$l=8$ & $0.001137050$ & $k=10^3$ & $5.999256\cdot 10^{-7}$ \\ 
\hline
$l=9$ & $0.001425472$ & $k=10^6$ & $6.000149\cdot 10^{-13}$ \\ 
\hline
\hline
Parameters & $\alpha$ & Parameters & $\alpha$ \\ 
$S=-2c=k/80=1$ & & $S=-10c=k/40=1$ & \\
\hline
$l=4$ & $0.001562306$ & $l=4$ & $0.001263604$ \\ 
\hline
$l=5$ & $0.002347145$ & $l=5$ & $0.001907838$ \\ 
\hline
$l=6$ & $0.003292258$ & $l=6$ & $0.002692482$ \\ 
\hline
$l=7$ & $0.004399501$ & $l=7$ & $0.003624653$ \\ 
\hline
\end{tabular}
\end{tiny}
\end{table}

The effect of a dark matter component in the spectra of a Schwarzschild-dS like black hole is that of increasing the imaginary part when $k>0$ and lowering it when $k<0$. In table \ref{table:tab1} we list the quasinormal modes of different black holes calculated with different methods, showing good agreement between the obtained data: the highest deviation in the results is around $0.2\%$ when $k=-0.05$.

With fixed parameters $S=l=-c/10=1$, if we change $k$ from its critical point up to $k\sim 0.5$, the potential reaches the regime of negative regions between horizons, affecting the relation of $\omega_R$ and $\omega_I$ to $k$. Going to higher $k$, the ringing phase vanishes and an exponential decay takes place whose behaviour is contrary to the previous case: the higher the $k$, the smaller the $\alpha$. In particular, when $k>>S,c,l$, we have a scaling between the decay coefficient and the parameters of the black holes as
\be
\label{esc1}
\alpha = l(l+1) \frac{c}{k^2}.
\ee
In  table \ref{table:tab3} we present a quasinormal spectrum with fixed black hole parameters under different $l$. The effect of increasing $l$ on the spectrum is very mild in the imaginary part, (decreasing  very slowly) but more robust in $\omega_R$ (increasing the frequency).

\begin{table}[ht]
\caption{Different $l$ for fixed black hole parameters.}
\label{table:tab3}
\begin{tabular}{|c|c|c|} 
\hline
$l$ & Characteristic integration &  $6^{th}$-order JWKB  \\
\hline
$3$ & $1.3454-0.19439i$ & $1.3451-0.19452i$ \\
\hline
$4$ & $1.7281-0.19412i$ & $1.7279- "0.19417i$ \\
\hline
$5$ & $2.1112-0.19392i$ & $2.1110-0.19399i$ \\
\hline
$6$ & $2.4945-0.19379i$ & $2.4941-0.19389i$ \\
\hline
$7$ & $2.8780-0.19370i$ & $2.8774 - 0.19382i$ \\
\hline
$8$ & $3.2616-0.19362i$ & $3.2607-0.19378i$ \\
\hline
$9$ & $3.6453-0.19355i$ & $3.6440-0.19375i$ \\
\hline
\end{tabular}
\end{table}

\section{Schwarzschild-AdS-dM Black Hole \label{ssdsddm}}

\begin{figure}[h]
\caption{Limit functions $X_\pm$ for a three horizons causal structure AdS black hole. The yellow region represents possible values for dark energy, $c$.}
\resizebox{\linewidth}{!}{\includegraphics{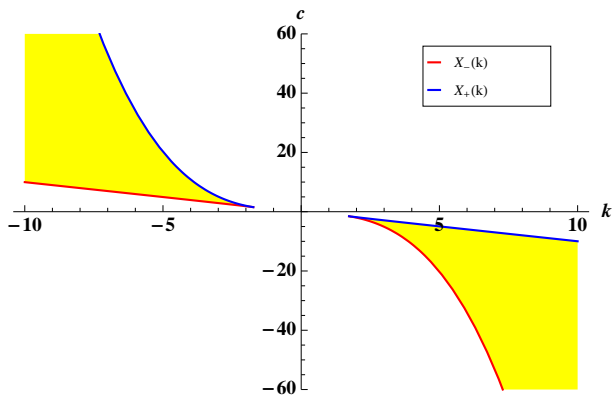}\label{f1}}
\end{figure}

Out anti-de Sitter-Schwarzschild-type black hole with a dark matter component has the metric coefficient $g_{tt}$ given by 
\be
\nonumber
-g_{tt}=1 - \frac{S}{r} + (cr+k)^2=\frac{c^2 \left( r^3 + \frac{2k}{c}r^2+ \frac{k^2+1}{c^2}r-\frac{S}{c^2} \right)}{r}
\\
\label{e2}
\ee
Considering the three possible roots for $g_{tt}=0$, we have $r_1r_2r_3=S>0$ such that we must have at least one positive root for the horizon equation, even if the two others roots are not real; as a consequence, two different status arise:
\vskip2ex
{\it (i) Three Horizons-spacetime}: this shall be the case when the following conditions hold,
\\ 
(A) $|k|>\sqrt{3}$; 
\\ 
(B) $kc<0$; 
\\
(C) $X_-< c <X_+$, \\
with $X_\pm= X_\pm (S,k)= \left( -\frac{2k^3+18k}{27S} \pm \frac{2}{27S}\sqrt{(k^2-3)^3}  \right)$. Condition (C) may still be written as $H_+< S <H_-$,  with $H_\pm= H_\pm (c,k)= \frac{2}{27c}\sqrt{(k^2-3 )^3} \pm \frac{2k^3+18k}{27c}$. We shall assume $r_1<r_2<r_3$. In figure \ref{f1} we see a plot of the functions $X_\pm$ with $S=1$ and the possible range for cosmological constant in case $(i)$. Besides conditions (A) to (C), the three horizons status can be assured for every pair $(S,c)$, when the inequality $k_-<k<k_+$ holds, being $k_\pm$ the solution for the equation $X_\pm (k_\pm)-c=0$.
In this case we shall have a dynamical Universe in regions $r<r_1$ and $r_2 < r < r_3$, and a static Universe in regions $r_1 < r< r_2$ and $r>r_3$. For a three horizons solution, we study the evolution of the scalar field beyond the third horizon.

\vskip2ex
{\it (ii) One Horizon-spacetime}: should be the case if one of the conditions (A) to (C) does not hold. In this case we must have a static physical Universe after the event horizon $r_1$ and before this point, dynamical. The Klein-Gordon equation will be treated beyond $r=r_1$. 
\\

In both cases the only curvature singularity lies at $r=0$, thus encapsulated by one or more horizons.

\subsection{Effective potential}

The scalar field effective potential for the geometry reads
\be
\nonumber
V(r ) = \left[ 1 - \frac{S}{r} + (cr+k)^2 \right] \left[ \frac{S}{r^3} + 2c^2 + \frac{2ck}{r}+\frac{l(l+1)}{r^2}  \right].
\\
\label{b22}
\ee  
and it is always positive beyond the event horizon ($r_1$ or $r_3$), preventing the manifestation of instabilities in the quasinormal spectra. In the next figures \ref{f10} and \ref{f11} we see two plots for this potential, with three horizons and one horizon respectively.
\begin{figure}[h]
\caption{AdS-dM-Schwarzschild-DM black hole with 3 horizons. The first horizon reads $r_1 \sim 0.379$. The potential is never negative beyond $r_3=r_h$. The static region between $r_1$ and $r_3$ has a negative potential depending on $k,c$ and $l$.}
\resizebox{\linewidth}{!}{\includegraphics{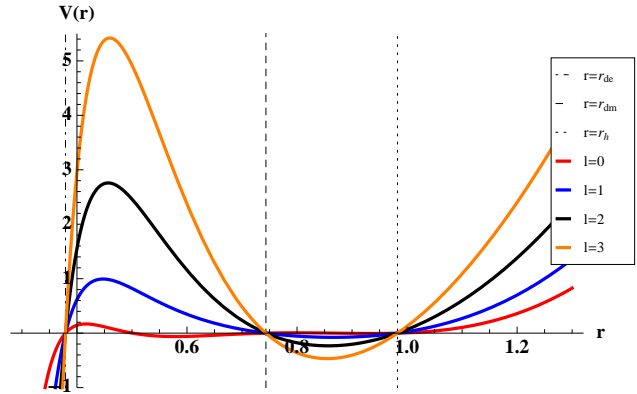}\label{f10}}
\end{figure}
\begin{figure}[h]
\caption{AdS-Schwarzschild-DM black hole with 1 horizon.}
\resizebox{\linewidth}{!}{\includegraphics{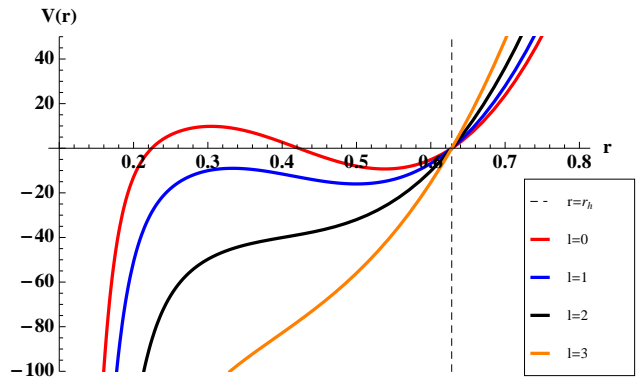}\label{f11}}
\end{figure}

In the first plot (figure \ref{f10}) the parameters of the geometry are given by $S=k/2=-c/1.9=1$. Other than the roots $(r_h,r_{dm},r_{de})$, we have $V(r)=0$ at a particular $r$ only when $l=0$, always before the event horizon such that it will not interfere with the scalar field evolution.


In the second plot we have $S=k/3=-c/6$, which presents the same qualitative behavior: $V(r)$ is zero only for the roots of $g_{tt}$ or when $l=0$ in two different points before the event horizon, again accommodating a mild evolution of fields outside the black hole.


Given the different potentials with three or one horizon, the qualitatively behavior for the scalar field is quite the same, which can be seen in both figures after $r=r_h$.

In what follows we present the scalar field evolution  obtained in null-coordinates, given proper boundary conditions with a variety of parameters and potentials described, in the range $r>r_h$. 

\subsection{Scalar field profile evolution}

As reported in the previous section, the evolution of the scalar field is obtained with different boundary conditions, proper to an AdS black hole. We analyze in separate the two different causal status, with one or three horizons.
\\

{\it One Horizon case}
\\

\begin{figure}[h]
\caption{Scalar field profile of AdS-Schwarzschild-DM black hole with 1 horizon. The used parameters read $S=c=1$ and $l=0$.}
\resizebox{\linewidth}{!}{\includegraphics{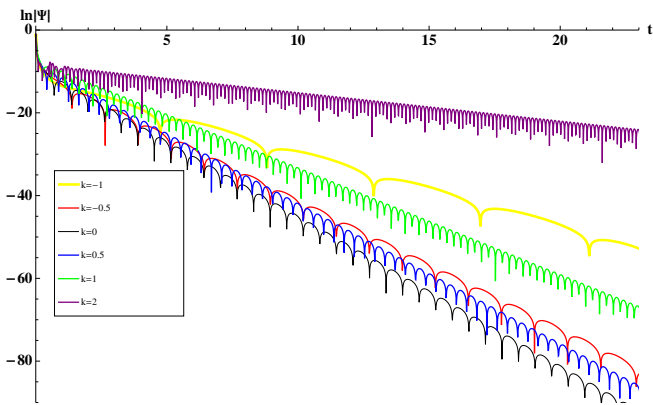}\label{f14}}
\end{figure}

In figure \ref{f14} we see some field profiles in time domain with different $k$ parameters.

In this figure we can see the influence of a dark matter term in an AdS-Schwarzschild solution: compared to the $k=0$ profiles, the dark matter black hole oscillates with a smaller damping, $\omega_I$, whatever the signal of $k$; the profiles with smallest damping are those with higher $|k|$. Concerning the frequency of oscillation $\omega_R$, it diminishes with the increment in $k$, for $k>0$; when $k<0$, the value of $\omega_R$ increases as $|k|$ increases up to a critical point, dropping off afterwards: taking $S=c=1$, $l=0$, $k_c \sim -1.2$. This turns out to be the point where the Hawking temperature of this black hole has its inflection point as well. 

For all the profiles analyzed in the range of parameters of this AdS single horizon black hole, the field has an ever oscillating record at late times, not showing a tail or exponential decay, which happens also in other  known solutions of AdS black holes.

Given the shape of the tortoise coordinates,  the computation of the signal is increasingly difficulted for high values of $k$: the function $\frac{d^2r}{dr_*^2}$ diverges for different points beyond the horizon, making the quasinormal ringing phase to appear in late times only, demanding great computational time to be computed.

\begin{figure}[h]
\caption{Scalar field profile of AdS-Schwarzschild-DM black hole with 1 horizon. The used parameters read $S=c=10k=1$}
\resizebox{\linewidth}{!}{\includegraphics{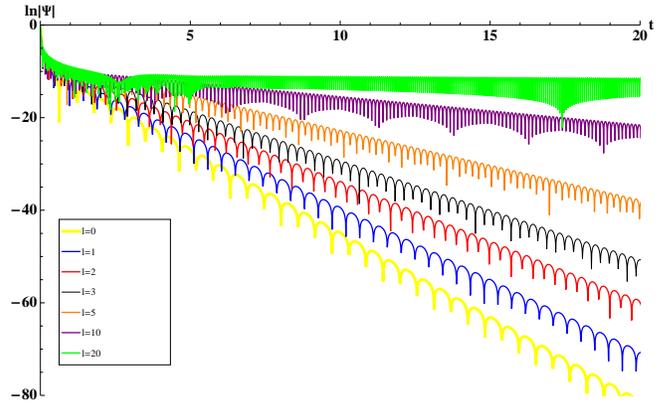}\label{f14b}}
\end{figure}

The same difficulty is found for the acquisition of the signal for high $l$: the higher the $k$ the higher the computational time to obtain profiles with high $l$. A plot with different $l$ profiles follows in figure \ref{f14b}.
In general $\omega$ displays smaller imaginary terms, and larger real ones, with increasing $l$, making it difficult to obtain modes with high $l$ for long time periods. This happens to be the same behavior found in \cite{horohub} without a dark matter component.

\begin{figure}[h]
\caption{Scalar field profile of AdS-Schwarzschild-DM black hole with 1 horizon. The used parameters read $c=k=1$ and $l=0$.}
\resizebox{\linewidth}{!}{\includegraphics{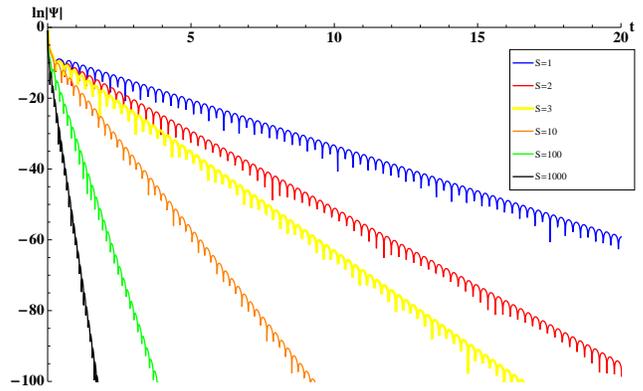}\label{f15}}
\end{figure}

The last behavior observed for the case of an one horizon black hole was to modify the event horizon increasing the Schwarzschild sphere, $S$, obtaining the field profiles displayed in figure \ref{f15}.

We observe an increment of $\omega$ concomitant to increasing $S$, a pattern fitting one of the peculiar interpretations of this metric under the AdS/CFT conjecture \cite{horohub}: $\omega_I$ is interpreted as the relaxation time in the Conformal Field Theory dual to an anti-de Sitter theory; we infer that $S$ keeps proportion to $\omega$ by the figure, but not in a linear correspondence. The relationship between $S$ and $\omega$ will be quantified in the next section where we calculate the quasi-normal modes of the black hole.
\\

{\it Three Horizons}
\\

In order to obtain a 3-horizon AdS black hole, conditions (A) to (C) must hold, that is, $c$ and $k$ have opposite sign. The case in which $c>0$ and $k<0$ is exactly the same inverting signals of $k$ and $c$. 

The range of parameters for which we can have a 3-horizon black hole is strictly limited; for example when $S=c/2=1$, we must have $ - 2.114 \leq k < -2$. Being $k^2>3$, high, the evolution of the field for small $l$ rapidly undergoes exponential decay, showing no quasinormal phase in most of the cases \footnote{If the quasinormal frequency $\omega_R$ is small, it is not possible to see the oscillation in the spectrum for small $l$, given its short span.}.

Now, for small $(k,c)$ in the 3-horizon spacetime, we may have a quasinormal ringing phase observable only for high multipole number. 

\begin{figure}[h]
\caption{Scalar field profile of AdS-dM-Schwarzschild black hole with 1 horizon. The used parameters read $S=c/5=1$ and $l=0$.}
\resizebox{\linewidth}{!}{\includegraphics{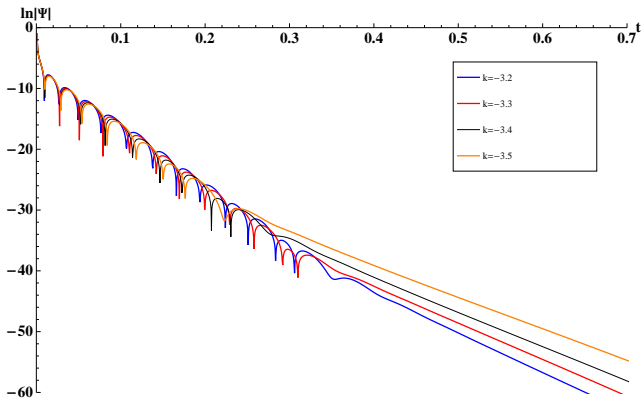}\label{f155}}
\end{figure}

In figure \ref{f155} we see profiles of field evolution in time domain for different $k$: taking $k$ in the range $k_-\sim 3.171 < k < k_+ \sim 3.560$, we can see that the field rapidly goes into  exponential decay when $l=20$. The evolution with $l\leq 10$ does not appear as a quasinormal ringing phase at all, and the field, after a initial burst decays exponentially. 

The peculiar behavior of  $|k|$ proportional to $\omega_I$ is more subtle to draw in the 3 horizon case: the quasinormal ringing phase happening in a short range rise the uncertainty with Prony calculations; however, the long time behavior shows the same peculiar behavior: increasing $|k|$ produces less damping of the exponential decay. The fact was tested for a wider range of values, the same pattern being found.

\subsection{Quasinormal modes}

Given the profiles in $t$-domain as shown in the previous subsection, we apply the Prony method (already mentioned) or a simple linear regression to obtain the exponential decay in the case of 3-horizon black hole. For the procedure we suppose a solution of kind $\Psi (t) \rightarrow e^{-i\omega t}$, being $\omega = \omega_R - i \omega_I$. Other than this technique, it is possible to use the method developed in \cite{horohub} (series expansion around $r_h$) to extract the frequencies, whose convergence is poorly obtained in our case of AdS dark matter black hole for small parameters of $S,k$ and $l$. In the case without dark matter ($k=0$), we tested the null coordinates field evolution along with the Prony method and obtained good agreement with the Frobenius expansion, as seen in table \ref{table:tab4}.
\begin{table}[h]
\caption{Quasinormal modes for AdS-Schwarzschild black hole with characteristic integration/prony and Frobenius methods. The parameters taken are $c=1$ and $l=0$. \label{table:tab4}}
\begin{tabular}{|c|c|c|} 
\hline
$r_h$ & Frobenius & Prony  \\
\hline
$0.6$ & $2.4316-1.5797i$ & $2.4293-1.5798i$  \\
\hline
$0.8$ & $2.5878-2.1304i$ & $2.5878-2.1303i$  \\
\hline
$1.0$ & $2.7982-2.6712i$ & $2.7982-2.6711i$  \\
\hline
\end{tabular}
\end{table}

For the black hole with one horizon we applied the Prony method and obtained the quasi-frequencies, first with fixed $c$, $S$ and $l$ and varying the dark matter component $k$, then afterward fixing $k$ and varying $l$. The results follow in table \ref{table:tab5}.

\begin{table}[h]
\caption{Quasinormal modes for AdS-Schwarzschild-DM black hole with 
$c=S=1$, $l=0$ for varying $k$, and $S=c=k/10=1$ for varying $l$.\label{table:tab5}}
\begin{tabular}{|c|c|c|c|} 
\hline
$k$ & $\omega$ & $l$ & $\omega$ \\
\hline
$-1.4$ & $0.3639471-0.1373848i$ & 0 & $2.772190-1.812461i$ \\
\hline
$-1.2$ & $0.2466027-0.4820482i$ & 1 & $3.407416-1.556096i$ \\
\hline
$-1$ & $0.3860446-0.8821291i$ & 2 & $4.297669-1.289525i$ \\
\hline
$-0.5$ & $1.247400-1.556081i$ & 3 & $5.276812-1.072440i$ \\
\hline
$0$ & $2.487681-1.808024i$ & 4 & $6.293203-0.8976747i$  \\
\hline
$0.5$ & $4.035366-1.692266i$ & 5 & $7.327595-0.7542458i$ \\
\hline
$1$ & $5.935457-1.289801i$ & 7 & $9.425730-0.5333276i$ \\
\hline
$1.5$ & $8.246449-0.7646561i$ & 10 & $12.59145-0.3065000i$ \\
\hline
$2$ & $10.91329-0.3435674i$ & 20 & $23.10316-0.01823809i$ \\
\hline
\end{tabular}
\end{table}
We may notice a peculiar feature for the AdS dark matter black hole: the highest damping modes are those with high $|k|$, which is true also for other range of values calculated. This is the contribution of a dark matter term to the AdS-Schwarzschild black hole: it diminishes the slope of the field evolution in relation to black holes with no dark matter term. As of the dependence in angular momentum, the field presents the same behavior as in an AdS-Schwarzschild black hole: $\omega_R$ increases at the same time $\omega_I$ decreases for increasing $l$. 

With the single horizon black hole we now check the scaling between the black hole event horizon temperature and the quasinormal spectrum in both real and in imaginary parts. In table \ref{table:tab6} we list a group of calculated frequencies for different $S$, $r_h$ and $T_H$ with constant $c, k$ and $l$.

\begin{table}[h]
\caption{Quasinormal modes for AdS-Schwarzschild-DM black hole with 
$c=k=1$, $l=0$.\label{table:tab6}}
\begin{tabular}{|c|c|c|c|} 
\hline
$S$ & $r_h$ & $T_H$ & $\omega$ \\
\hline
$1$ & $0.35321$ & 0.8532284 & $5.935457-1.289800i$ \\
\hline
$2$ & $0.57474$ & 0.7324346 & $6.057333-2.187124i$ \\
\hline
$3$ & $0.74296$ & 0.7098959 & $6.234276-2.796013i$ \\
\hline
$10$ & $1.43906$ & 0.7724561 & $7.226347-5.013375i$ \\
\hline
$100$ & $3.938598$ & 1.298990 & $11.53555-12.02626i$  \\
\hline
$1000$ & $9.313585$ & 2.558853 & $21.32894-26.48288i$ \\
\hline
$10^6$ & $99.33114$ & 24.03347 & $187.7282-266.3216i$ \\
\hline
$10^{12}$ & $999.3331$ & 238.8917 & $1852.589-2663.253i$ \\
\hline
$10^{15}$ & $99999.33$ & 23873.40 & $184988.2 - 266324.1i$ \\
\hline
\end{tabular}
\end{table}

There is no linear proportion between the temperature/horizon of the hole and the quasi-frequency (real or imaginary part) as found in the standard AdS-Schwarzschild case \cite{horohub}; however there is a proportional relation between $r_h$ ($T_H$) and $\omega$ for ˜higher values of horizon™ $r_h$; increasing $r_h$ causes $\omega$ to increase as well. In particular the slope of $\omega_I / T_H$ approaches 11.16 and the slope of $\omega_I / r_h$ approaches 2.66 for high values of $S$, the same factor found in \cite{horohub} for high $r_h$.

The last part of the results for the scalar field evolution concerns the case of a 3-horizon AdS-DM black hole, which shows for small $l$ no quasinormal modes at all. For high $l$, a quasinormal ringing phase seems to be formed, but going for high values of $k$ the ringing diminishes and eventually vanishes. The field profiles in the 3-horizon case decays exponentially very rapidly at a higher rate for smaller $|k|$. This fact resembles the behavior obtained in the previous case for which the higher the $|k|$, the smaller the $\omega_I$. An example of field decay evolution in time follows in table \ref{table:tab7}.

\begin{table}[h]
\caption{Field coefficient decay of a 3-horizon AdS-Schwarzschild-DM spacetime. The parameters taken are $S=c/5=1$ and $l=0$.\label{table:tab7}}
\begin{tabular}{|c|c|} 
\hline
$k$ & decay  \\
\hline
$-3.2$ & $64.27743i$ \\
\hline
$-3.3$ & $61.75985i$  \\
\hline
$-3.4$ & $56.42453i$ \\
\hline
$-3.5$ & $51.92062i$ \\
\hline
\end{tabular}
\end{table}

\section{Final Remarks}
\label{conc}
We have computed, by use of three well established methods, the quasinormal modes of a Schwarzschild-de Sitter (and Schwarzschild-anti de Sitter) metric with presence of dark matter and cosmological constant terms, deduced by direct use of the Nash formalism for embedding a brane solution into a 5-dimensional constant cuvature bulk. We have compared the obtained modes vis a vis the most similar counterpart metric in standard General Relativity. 

The remarkable feature of Nash's formalism is to provide a differential and continuous embedding scheme without use of arbitrary mathematical assumptions for the brane junctions. Its rendering of a unified geometric picture for both dark energy and dark matter behavior is of notice, but given the many possible embedding formalisms in the literature, there is the need to produce observable and differentiating predictions. A few distinctions follow from our results:

\textit{(i)} Our $l=0$ SdS-DM case produces no relevant differential signature compared to GR, but for all the other angular momenta the damping is stronger ($k>0$) or weaker ($k<a$) than standard GR. Given different galaxy or cluster rotation curves profiles, providing fittings for the values of the dark matter component $k$, we conjecture there should be a clear correlation to the ringdown of binary black hole mergings located in the same regions. The analysis of such tail effects on gravitational wave signals is a work in progress.

\textit{(ii)} Though of less relevance for astrophysics, our AdS-Schwarzschild-Dark Matter case bears relevance to AdS/CFT scenarios \cite{horohub}, showing the scaling of the finite CFT temperature with the AdS ringing modes is more nuanced in the presence of dark matter components. Given that, in our formalism, both dark matter and dark energy have geometrical origin, i.e. correspond to different irreducible geometrical degrees of freedom, it is natural to expect their CFT dual will be changed. Given the $k$ term's influence falls with greater temperature, it resembles the behavior of holographic AdS/CFT superconductors in rotating spacetimes \cite{sonner,lin}, whereas the extra degree of freedom from the black hole angular momentum is here substituted by the $k$ degree of freedom.\\

Further comparison with different braneworld formalisms shall enlighten our knowledge of the role of extra dimensions in providing observable signatures of new physics, for both astrophysical and AdS/CFT phenomenological analyses.

\begin{acknowledgments}
The authors thank CAPES and CNPq for financial support.

\end{acknowledgments}

\end{document}